\documentclass[journal=jpccck,manuscript=article]{achemso}
\usepackage{graphicx,subcaption,amsmath,gensymb,esvect,xcolor,xr}

\setkeys{acs}{maxauthors=30, etalmode=truncate}

   \makeatletter
\newcommand*{\addFileDependency}[1]{
  \typeout{(#1)}
  \@addtofilelist{#1}
  \IfFileExists{#1}{}{\typeout{No file #1.}}
}
\makeatother

\usepackage[version=3]{mhchem}

\newenvironment{competing interests}
{
  \par\vspace{\baselineskip} \noindent
  \begin{Large}\textbf{Competing Interests}\end{Large} 
  \par \noindent\ignorespaces
}

\newenvironment{data availability}
{
  \par\vspace{\baselineskip}\noindent
  \begin{Large}\textbf{Data Availability}\end{Large}
  \par \noindent\ignorespaces
}

\newenvironment{author contribution}
{
  \par\vspace{\baselineskip}\noindent
  \begin{Large}{\textbf{Author Contribution}} \end{Large}
  \par \noindent\ignorespaces
}

\author{Shahid Sattar}
\email{shahid.sattar@ltu.se}
\author{J. Andreas Larsson}
\email{andreas.1.larsson@ltu.se}
\affiliation{Applied Physics, Division of Materials Science, Department of Engineering Sciences and Mathematics,
Lule\aa\,University of Technology, Lule\aa\, SE-97187, Sweden}
\phone{+46 (0)920 491930}

\title{Rashba Effect and Raman Spectra of Tl$_2$O/PtS$_2$ Heterostructure}

\keywords{platinum disulphide, thallium oxide, metal oxide monolayer, heterostructure, Rashba, Raman}

\begin{document}

\begin{tocentry}
\includegraphics{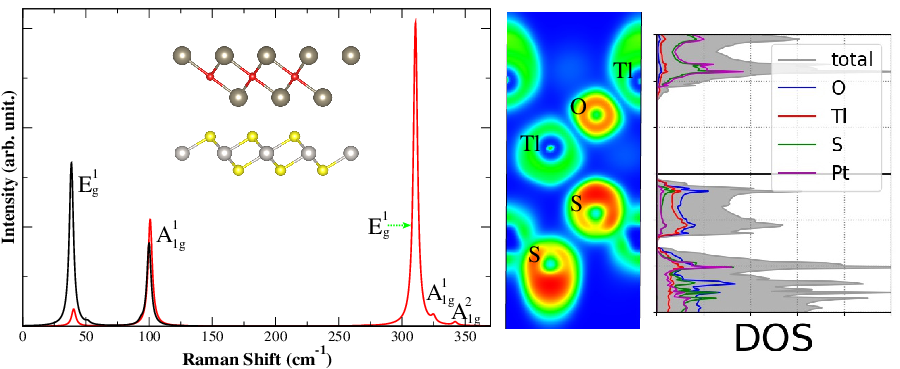}
\end{tocentry}

\begin{abstract}

The possibility to achieve charge-to-spin conversion via Rashba spin-orbit effects provide stimulating opportunities toward the development of nanoscale spintronics. Here we use first-principles calculations to study the electronic and spintronic properties of Tl$_2$O/PtS$_2$ heterostructure, for which we have confirmed the dynamical stability by its positive phonon frequencies. An unexpectedly high binding energy of -0.38 eV per unit cell depicts strong interlayer interactions between Tl$_2$O and PtS$_2$. Interestingly, we discover Rashba spin-splitting's (with large $\alpha_R$ value) in the valence band of Tl$_2$O stemming from interfacial spin-orbit effects caused by PtS$_2$. The role of van der Waals binding on the orbital rearrangements has been studied using electron localization function and atomic orbital projections, which explains in detail the electronic dispersion near the Fermi level. Moreover, we explain the distinct band structure alignment in momentum space but separation in real space of Tl$_2$O/PtS$_2$ heterostructure. Since 2D Tl$_2$O still awaits experimental confirmation, we calculate, for the first time, the Raman spectra of pristine Tl$_2$O and the Tl$_2$O/PtS$_2$ heterostructure and discuss peak positions corresponding to vibrational modes of the atoms. These findings offer a promising avenue to explore spin physics for potential spintronics applications via 2D heterostructures.  
\end{abstract}

\section{Introduction}

Spin-orbit effects in two-dimensional (2D) materials are of central importance for designing next-generation spintronic, valleytronic and spin-logic memory devices. Transition-metal dichalcogenides (TMDCs) such as MoS$_2$, in their pristine form and in proximity to other 2D materials, have been extensively utilized in such applications owing to their large spin-orbit strength and other promising features \cite{yan2016two,radisavljevic2011single,zeng2012valley,ghiasi2017large}. In this context, the Rashba spin-orbit effects (or simply Rashba effect) are of particular interest because it enables charge-to-spin conversion in a non-magnetic material by lifting spin-degeneracy along the momentum-axis without the need of an external magnetic field \cite{bychkov1984properties,manchon2015new}. The effect is observed experimentally in a variety of materials (e.g., in metal surfaces \cite{lashell1996spin,koroteev2004strong,kimura2010strong,su2017selective,hirahara2006role,hochstrasser2002spin,dil2008rashba,varykhalov2012ir}, bulk materials \cite{ishizaka2011,di2013electric,sakano2013strongly,feng2019rashba}, perovskites \cite{niesner2016giant,niesner2018structural,zhai2017giant}, topological insulators \cite{zhu2011rashba,king2011large,zhou2014engineering}, oxides \cite{lin2019interface} and 2D materials \cite{yao2017direct,khokhriakov2020gate}) with a continued surge to achieve better control of the spin degree of freedom of electrons in new materials. Moreover, the possibility to build lateral and vertical 2D heterostructures without the constraints of lattice matching further broadens the scope of new discoveries and realization of novel phenomenon.

Monolayer Tl$_2$O is a recently proposed 2D metal-oxide semiconductor with cleavage energy comparable to well-known TMDCs \cite{ma2017single}. Several theoretical studies highlighted its potential in catalysis \cite{li2020theoretical}, valleytronics \cite{xu2019nonmetal}, and especially in thermoelectrics \cite{wang2019theoretical,huang2019layered} due to the ultralow lattice thermal conductivity \cite{sajjad2019ultralow}. Bulk Tl$_2$O crystallizes in the 1T-phase with its cousin polytype also existing in 2H-phase albeit higher in energy \cite{ma2018conduction}
thus favoring the synthesis of the former due to energetic reasons. While 2D Tl$_2$O still awaits experimental realization, a 2D thallene was recently fabricated on NiSi$_2$/Si(111) substrate experiencing a strong tensile strain due to the large lattice mismatch \cite{gruznev2020thallene}. It is therefore extremely critical to select an appropriate material capable to host Tl$_2$O in its most stable form (without causing detrimental effects to its structure and properties) and permits usage in advanced applications. 

In the present study, we use electronic structure theory calculations to demonstrate that 1T-Tl$_2$O in heterostructure with the lattice-matched PtS$_2$ has multiple advantages: 1) The heterostructure is dynamically stable as confirmed by the phonon band structure showing positive frequencies throughout the Brillouin zone. 2) An unusually high binding energy of -0.38 eV (per unit cell) depicts strong attachment between Tl$_2$O and PtS$_2$ with structural features of both materials largely remain intact. 3) We observe giant Rashba spin-splittings with large Rashba ($\alpha_R$)-parameter in the Tl$_2$O/PtS$_2$ heterostructure compared to previously studied heterostrucures of 2D materials. These findings are further confirmed by spin-texture plots with detailed analysis of the electron localization function (ELF) and electronic density of states (DOS) in connection with the interfacial spin-orbit effects. We also note that for a small biaxial tensile strain, the characteristic Rashba splittings near the Fermi level is maintained. Furthermore, to expedite a quick experimental confirmation of our work, we present for the first time the Raman spectra of pristine Tl$_2$O and the Tl$_2$O/PtS$_2$ heterostructure. Owing to the importance of spin generation and charge-to-spin conversion, our results provide crucial insights into Tl$_2$O heterostructures that offers rich spin-valley physics and potential within spintronics applications. 

\section{Computational method}

First-principles calculations have been performed using density functional theory (DFT) with projector augmented waves\,\cite{paw1,paw2} as implemented in the Vienna Ab-initio Simulation Package \cite{vasp}. We used the generalized gradient approximation in the Perdew-Burke-Ernzerhof parametrization to describe
the exchange-correlation effects, with a plane wave cutoff energy set to 450 eV. The van der Waals interactions have been taken into account using the DFT-D3 method \cite{grimme2010consistent}. A gamma-centered $12\times 12\times 1$ k-mesh was employed for the structural relaxation and in the self-consistent calculations, Brillouin zone integration was performed using a dense $20\times20\times 1$ k-mesh. Due to the involvement of heavy elements, spin-orbit effects were included in the band structure and density-of-states (DOS) calculations. To calculate the phonon band structure, we used the Phonopy package \cite{togo2015first} using a $4\times 4\times 1$ supercell of Tl$_2$O/PtS$_2$ heterostructure with a $4\times 4\times 1$ k-mesh. Moreover, 2D spin-textures were computed by setting up a 2D k-mesh ($k_x\times k_y:30\times30$) centered at the gamma-point ($k_z=0$). For the iterative solution of the Kohn-Sham equations, we achieved an energy convergence of $10^{-6}$ eV and a force convergence of $10^{-3}$ eV/\AA\, in our calculations. To avoid out-of-plane periodic image interactions, we have also used a 15\,\AA\,thick layer of vacuum. The Raman spectra of Tl$_2$O and the Tl$_2$O/PtS$_2$ heterostructure were computed by calculating the derivative of the macroscopic dielectric tensor with respect to the mode coordinates using VASP and Phonopy \cite{togo2015first} assisted \textit{raman-sc} python package \cite{vasp-raman-py}. The results were plotted using the Pyprocar \cite{pyprocar} and Matplotlib software packages \cite{matplotlib}.

\begin{figure}[!t]
\includegraphics[width=1.0\textwidth]{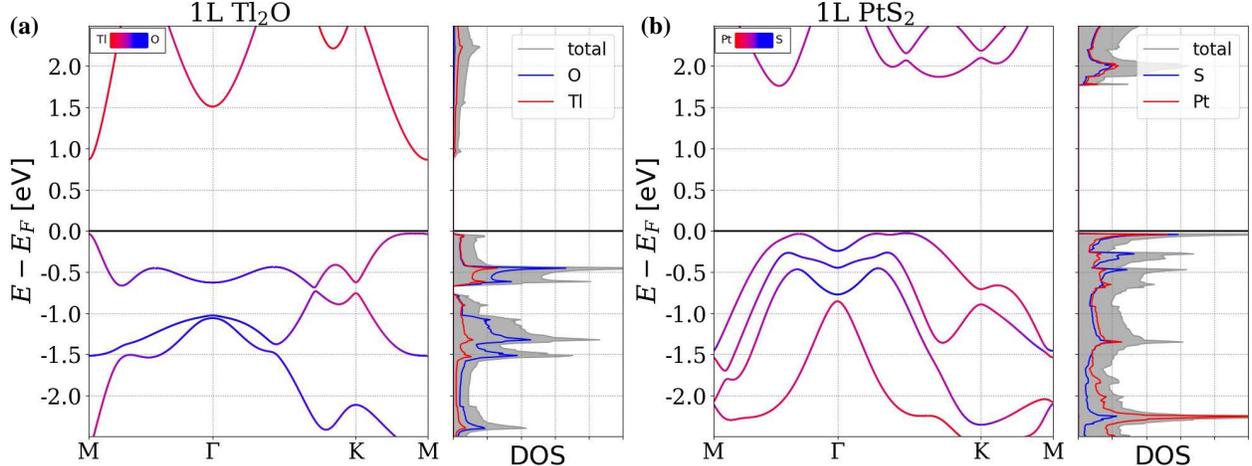}
\caption{Atom-projected electronic band structure and density-of-states (DOS) of monolayer (1L) (a) Tl$_2$O and (b) PtS$_2$, respectively. Spin-orbit coupling is included in both cases.}
\label{fig:fig1}
\end{figure}

\section{Results and Discussion}

Monolayer Tl$_2$O and PtS$_2$ both adopts 1T-phase trigonal prismatic geometry with an oxygen (O)(or platinum (Pt)) atom covalently bonded to two thallium (Tl) (or sulphur (S)) atoms, respectively. The optimized lattice parameter of 3.56\,\AA\,(3.58\,\AA) for Tl$_2$O (PtS$_2$) and band gaps of 0.90 eV (1.80 eV), respectively, are in close agreement to the existing reports \cite{sajjad2019ultralow,zhao2016extraordinarily}. Because of the lattice matched crystal structures, the possibility of epitaxial growth of Tl$_2$O on PtS$_2$ is anticipated. The atom-projected electronic band structure and density of states (DOS) for pristine monolayer Tl$_2$O and PtS$_2$ are shown in Figure \ref{fig:fig1}(a-b). Looking at the valence band of Tl$_2$O, it consists of mixed Tl and O atomic states whereas the conduction band mainly has Tl contributions, as depicted in the atom-projected DOS shown in Figure \ref{fig:fig1}a. On the other hand, PtS$_2$ show mixed contribution of Pt and S atomic states in both valence and conduction bands, as shown in Figure \ref{fig:fig1}b.

\begin{figure}[!t]
\includegraphics[width=0.99\textwidth]{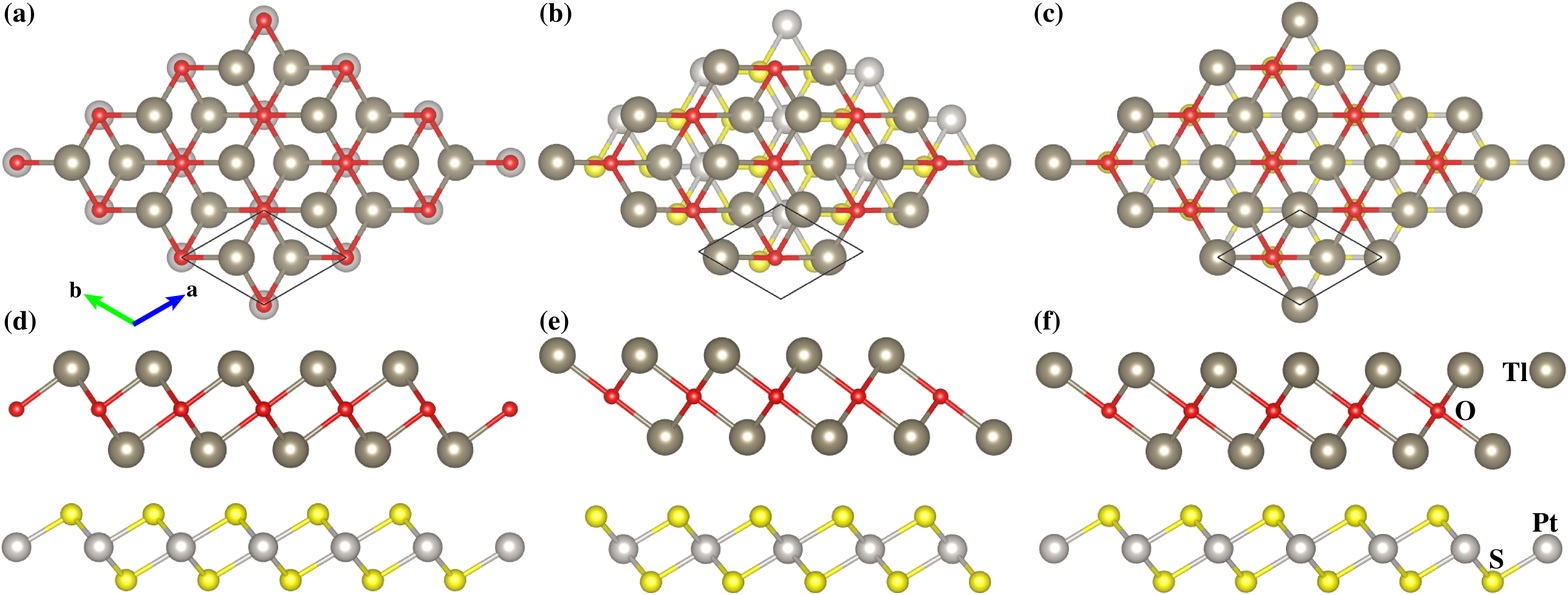}
\caption{(a-c) Top views and (d-f) side views of Tl$_2$O/PtS$_2$ heterostructure in different lateral stacking configurations. Unit cell is shown as black rectangle for each case.}
\label{fig:fig2}
\end{figure}

In order to built Tl$_2$O/PtS$_2$ heterostructures, we considered different lateral stackings by selecting multiple sites on PtS$_2$. Figure \ref{fig:fig2} shows top and side views of three such stacking configurations. In Figure \ref{fig:fig2}a/d the Tl atom lie exactly on top of S atoms. To scan different possible stackings, we traversed along $a-$ and $b-$axes, and also considered the inversion of Tl$_2$O as shown in Figure \ref{fig:fig2}(b/e). The stacking where O atoms lie exactly on top of top S atoms of PtS$_2$ turns out to be the minimum energy configuration as shown in Figure \ref{fig:fig2}(c/f). It is pertinent to mention that there exists also energetically degenerate configurations owing to small atomic displacements after structural relaxation. The configuration of Figure \ref{fig:fig2}(c/f) is, for comparison, 128 meV lower in energy compared to that in Figure \ref{fig:fig2}(a/d) and thus it is used in the rest of the calculations.

\begin{figure}[!htb]
\includegraphics[width=0.5\textwidth]{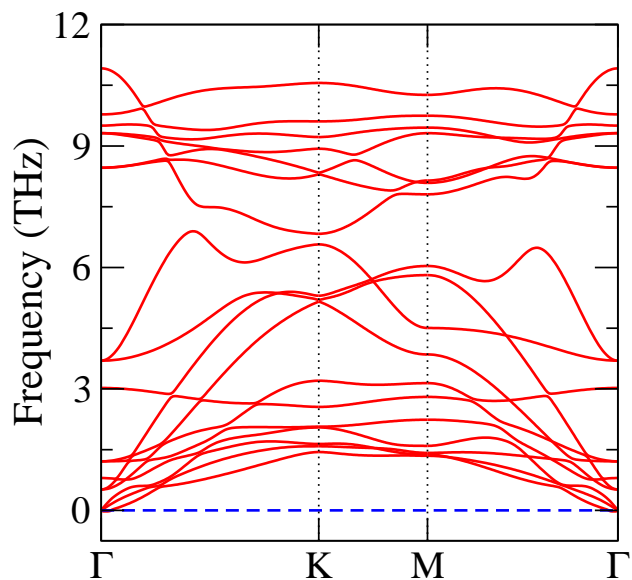}
\caption{Phonon band structure of the Tl$_2$O/PtS$_2$ heterostructure.}
\label{fig:fig3}
\end{figure}

We first checked the dynamical stability of the Tl$_2$O/PtS$_2$ heterostructure by calculating the phonon band structure. As displayed in Figure \ref{fig:fig3}, the phonon spectra for the minimum energy configuration does not show any imaginary frequencies, predicting the heterostructure to be stable. Moreover, we find coupling between the acoustic and optical phonons due to the strong interlayer interactions between the constituent systems as discussed below. 

To calculate the extent of binding between the constituent systems, we calculate the binding energy (per unit cell as marked in Figure \ref{fig:fig2}) through equation (\ref{equation:equation1}),
\begin{equation}
E_b=E({\rm Tl_2O/\rm PtS_2})-E({\rm Tl_2O})-E({\rm PtS_2}), \label{equation:equation1}
\end{equation}
where $E({\rm Tl_2O/\rm PtS_2})$ is the total energy of the heterostructure, $E({\rm Tl_2O})$ is the total energy of pristine Tl$_2$O, and $E({\rm PtS_2})$ is the total energy of pristine PtS$_2$. The obtained binding energy of $-0.38$ eV (per unit cell) shows unusually strong binding between Tl$_2$O and PtS$_2$, compared to similar vdW heterostructures \cite{nguyen2019controlling,alpha1}. An interlayer distance of $2.85\,\AA$ also supports this observation. As previous studies highlighted the importance of testing different van der Waals density functionals in this realm \cite{zhu2015silicene}, we also performed structural relaxation, and subsequently calculated the binding energy for the minimum energy configuration using the optB86b-vdW density functional \cite{optb86,optb86-2}. We obtain similar values using this functional and thus we can confidently categorize strong interlayer interactions in the Tl$_2$O/PtS$_2$ heterostructure.

\begin{figure}[!t]
\includegraphics[width=1.0\textwidth]{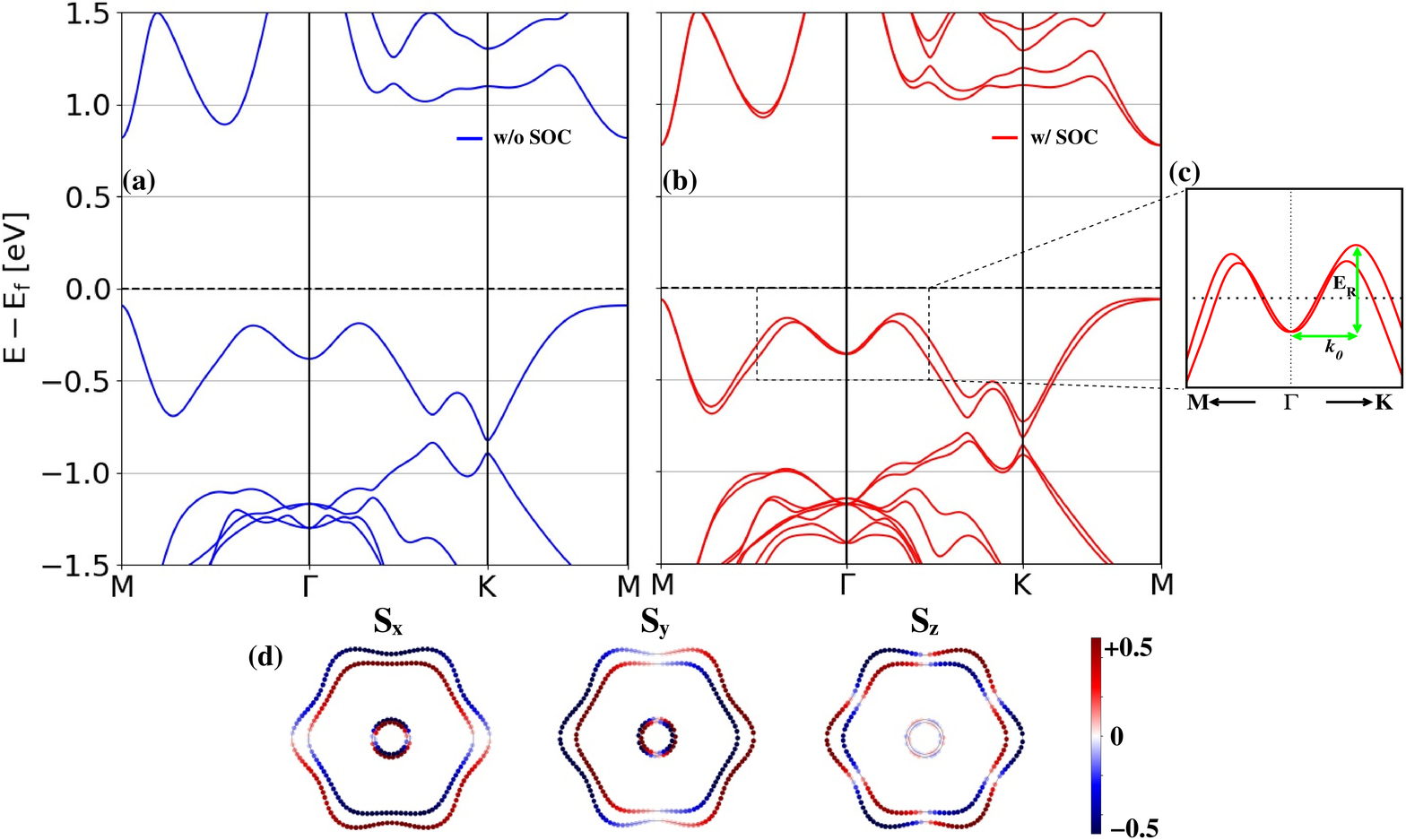}
\caption{Electronic band structure of Tl$_2$O/PtS$_2$ heterostructure (a) without (w/o) SOC and (b) with (w/) SOC, respectively. (c) Zoomed valence band region defining the generalized Rashba-energy $E_R$ and the momentum-offset $k_0$. (d) Fixed-energy spin-contour plots corresponding to dashed black line ($E=-0.30$ eV) in (c).}
\label{fig:fig4}
\end{figure}

Turning to the electronic and spintronic properties of the Tl$_2$O/PtS$_2$ heterostructure, we have calculated the electronic band structure given in Figure \ref{fig:fig4}(a-c). The constituent systems largely preserve their pristine band structures (cf. Figure \ref{fig:fig1}), but we observe a small reduction in the band gap of Tl$_2$O to a new value of 0.84 eV. Significant changes are however observed around the high symmetry gamma point near the Fermi level wherein the energy bands become more parabolic compared to pristine Tl$_2$O (cf. Figure \ref{fig:fig4}b and Figure \ref{fig:fig1}a). Moreover, we observe an upward shift of the energy bands at the high symmetry gamma point due to the orbital rearrangements caused by the interaction between Tl$_2$O and PtS$_2$. To investigate this further, we applied a small biaxial tensile strain up to $4\%$ and recalculated the electronic band structures (see Supplementary Figure S\ref{fig:fig1}). For the increased lattice constant in the strained heterostructure, the interlayer distance between Tl$_2$O and PtS$_2$ was slightly decreased in the structural relaxation (from $2.85\,\AA$ to $2.71\,\AA$). As a result, we observe a minuscule downward shift of the bands at the M-point and an upward shift at the gamma point. Most noticeably, inclusion of spin-orbit coupling show Rashba-type spin-splittings along the momentum-axis (i.e., spin-degenerate band splits into two parabolic bands with opposite polarities) around the high symmetry gamma-point (see Figure \ref{fig:fig4}b and zoomed region in Figure \ref{fig:fig4}c). To confirm this, we plot fixed-energy ($E=-0.30$ eV) contour plots of the spin-components ($s_x$, $s_y$ and $s_z$) by setting up a dense 2D k-mesh in the xy-plane (see Figure \ref{fig:fig4}d). The spin textures show clockwise and anticlockwise rotation of electron's spin in traversing from high symmetry M$\rightarrow\Gamma\rightarrow$K direction in the Brillouin zone validating these observations. We anticipate that the Rashba effect in Tl$_2$O/PtS$_2$ heterostructures can be electrically controlled by applying an external gate voltage. We employ the well-established Rashba Hamiltonian of a 2D-electron gas to describe the electronic dispersion according to equation (\ref{equation:equation2}), 
\begin{equation}
    H_R=\pm \frac{\hbar^2k_{\parallel}^2}{2m^*}+\alpha_R\vv{\sigma} . (\vv{k}_{\parallel}\times \vv{z}),
    \label{equation:equation2}
\end{equation}
in which $k_{\parallel}=(k_x,k_y,0)$ and $m^*$ is the in-plane momentum and effective mass of electron, respectively, $\vv{\sigma}$ is the vector of Pauli matrices and $\vv{z}$ is the out-of-plane unit vector. The Rashba parameter ($\alpha_R$) for a momentum-split parabolic-dispersion around the high symmetry gamma point, which represents the strength of the spin-orbit coupling, is approximated by $\alpha_R=2E_R/k_0$, whereas $E_R=\hbar^2k_{0}^2/2m^*$ and $k_0=m^*\alpha_R/\hbar^2$. Defining $E_R$ as the energy difference between the valence band maximum and the band crossing at the $\Gamma$-point and $k_0$ as the momentum offset, as seen in Figure \ref{fig:fig4}c. For the Tl$_2$O/PtS$_2$ heterostructure, we obtain large energy difference ($E_R=218$ meV) and momentum offset ($k_0=0.057\,\mathrm{\AA}^{-1}$) values thus resulting in the Rashba parameter $\alpha_R=7.65\,\text{eV}\,\mathrm{\AA}$. This value is higher than what has been found in the existing literature, which can be seen from our compilation in Table \ref{table:tables1}. Since the channel material in a typical spin field-effect transistor demands Rashba effect with large $\alpha_R$ values, Tl$_2$O in heterostructure with PtS$_2$ could be a potential candidate for building spintronics devices.

\begin{table}[!ht]
\caption{Rashba spin-splitting parameters of Tl$_2$O/PtS$_2$ heterostructure in comparison to previous works. $E_R$ is energy difference between the valence band maximum and band crossing at the $\Gamma$-point, $k_0$ is the momentum shift and $\alpha_R$ is the Rashba parameter.} 
\centering 
\begin{tabular}{c  c  c  c  r} 
\hline\hline 
Case & $E_R(\text{meV})$ & $k_0(\mathrm{\AA}^{-1})$ & $\alpha_R(\text{eV}\,\mathrm{\AA})$ & Reference \\ [0.5ex]
\hline
Tl$_2$O/PtS$_2$ Heterostructure   & 218 & 0.057 & 7.65 & \text{This work}\\[1ex]
GaSe/MoSe$_2$ Heterostructure     & 31  & 0.13  & 0.49 & \cite{alpha1}\\
PtSe$_2$/MoSe$_2$ Heterostructure & 150 & 0.23  & 1.30 & \cite{xiang2019tunable}\\
Bi/Ag(111) Surface Alloy          & 200 & 0.13  & 3.05 & \cite{ast2007giant}\\
Bulk BiTeI                        & 100 & 0.052 & 3.80 & \cite{ishizaka2011}\\
I-doped PtSe$_2$                  & 12.5 & 0.015 & 1.70 & \cite{absor2018strong}\\
LaOBiS$_2$                        & 38   & 0.025 & 3.04 & \cite{liu2013tunable}\\
MoSSe                             & 1.4  & 0.005 & 0.53 & \cite{li2017}\\[1ex] 
\hline 
\end{tabular}
\label{table:tables1} 
\end{table}

\begin{figure}[!htb]
\includegraphics[width=0.85\textwidth]{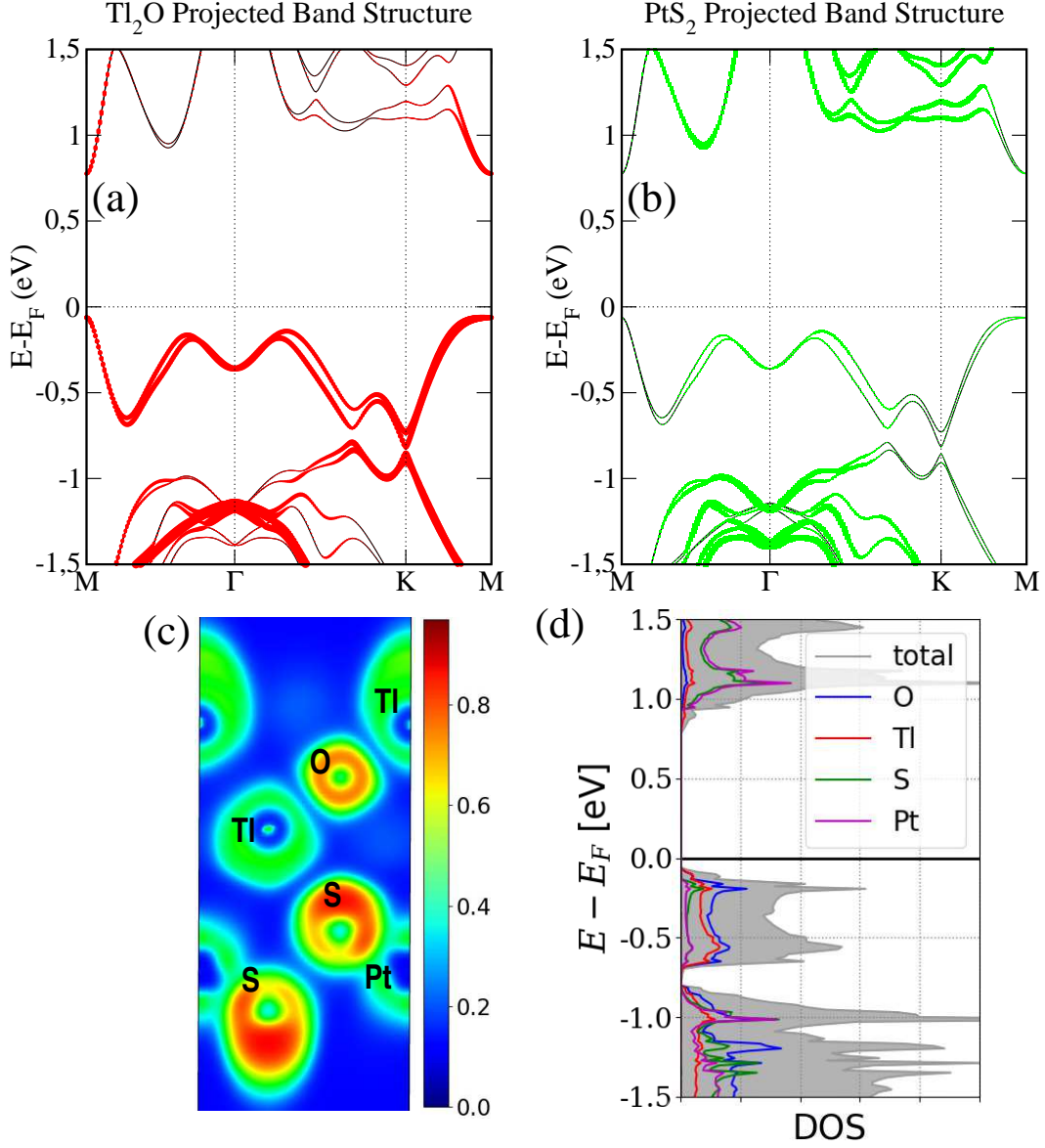}
\caption{Electronic band structure projected on (a) Tl$_2$O (b) PtS$_2$. (c) 2D Electron localization function corresponding to the (110)-surface passing through the heterostructure. (d) Total and atom-projected density of states (DOS).}
\label{fig:fig5}
\end{figure}

The importance of spin-orbit coupling in engineering the Rashba effect has been highlighted in several studies \cite{alpha1,xiang2019tunable}. To gain insights into the underlying mechanism in the Tl$_2$O/PtS$_2$ heterostructure responsible for this effect, we scrutinize the layer projected band structures in Figure \ref{fig:fig5}(a-b). Comparing the valence bands of the heterostructure with the individual monolayers in Figure \ref{fig:fig1}, we have found that the top valence band around the gamma point is dominated by Tl$_2$O but also has contributions from PtS$_2$. Moreover, looking at the atom-projected DOS given in Figure \ref{fig:fig5}d, despite PtS$_2$ electronic band dispersion to be present at an energy below $-0.80$ eV, we observe also small orbital contributions of Pt and S atoms in the energy window of $-0.25$ eV to $-0.75$ eV which depicts the same. This means that in the momentum space the two sides of the interface have bands contributing at the same energies around this point. Furthermore, the main contribution to the valence band close to the Fermi level and around the high symmetry gamma point is coming from $p-$orbitals, whereas there is mixed $p-$ and $d-$orbital contributions in the conduction band (see Supplementary Figure S\ref{fig:fig2}(a-b)). To further examine interlayer interaction, we also analyze the electron localization function (ELF) (see Figure \ref{fig:fig5}c), which shows remarkable contractions of the electron densities for the Tl and S atoms facing the interface (compare to the outer Tl and S atoms in the heterostructure) caused by the induced moments and Pauli repulsion due to the vdW binding. This is much more pronounced than the similar effect discussed for graphene/graphite in Ref. \cite{and1} because the vdW interaction between Tl$_2$O and PtS$_2$ in addition to dispersion contains dipole-dipole interactions as becomes evident when comparing the electronegativity for the interfacial atoms (Tl (1.62) and S (2.58)). This large perturbation of the PtS$_2$ bands around the gamma point gives rise to the interfacial spin-orbit coupling which in turn produces the Rashba effect. However, the ELF clearly shows that there are no bonds formed between Tl$_2$O and PtS$_2$, therefore we conclude that there is no hybridization between the two materials. Au contraire, our analysis of the physisorption between the materials show that their densities repel each other, which means that these states are separated in real space. Interface states that are separated in real space but locked together in momentum space due to strong physisorption seems to have been found in other 2D heterostructures \cite{xiang2019tunable}, but have not been evaluated properly. We suggest that designing strong physisorption in 2D heterostructures could be further explored to realize more and stronger spin physics phenomena.

\begin{figure}[!htb]
\includegraphics[width=0.85\textwidth]{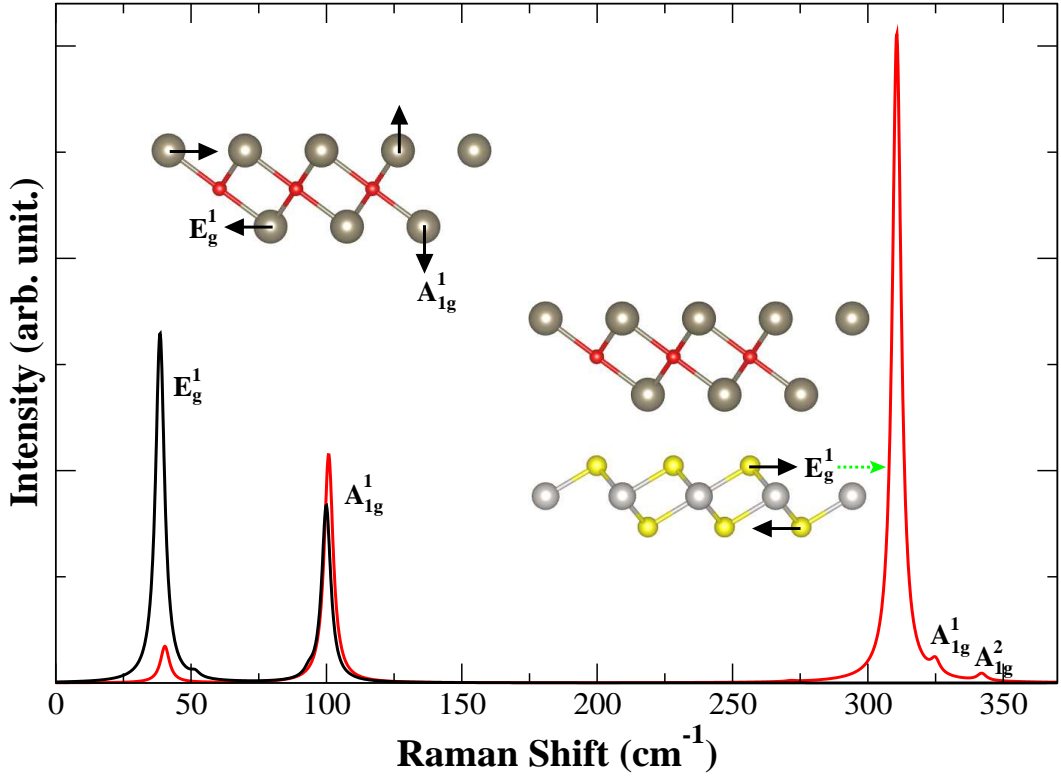}
\caption{Raman spectra of pristine Tl$_2$O (black) and Tl$_2$O/PtS$_2$ heterostructure (red).}
\label{fig:fig6}
\end{figure}

Finally, to further motivate experimental confirmation of our findings, we have also calculated the Raman spectra of pristine Tl$_2$O (in black) and the Tl$_2$O/PtS$_2$ heterostructure (in red) as shown in Figure \ref{fig:fig6}. For pristine Tl$_2$O, we observe two characteristic peaks corresponding to the E$_\text{g}^1$ and A$_{\text{1g}}^1$ modes at $40$ cm$^{-1}$ and $100$ cm$^{-1}$, respectively. The E$_\text{g}^1$ peak of high intensity correpsonds to the in-plane vibrational modes of Tl atoms within the 2D sheet, whereas the A$_{\text{1g}}^1$ peaks originate from the out-of-plane Tl atomic vibrations. In the Raman spectra of the Tl$_2$O/PtS$_2$ heterostructure, we observe additional peaks corresponding to PtS$_2$, i.e., the E$_\text{g}$ peak at $310$ cm$^{-1}$ and the minuscule A$_{\text{1g}}^1$ and A$_{\text{1g}}^2$ peaks at $325$ cm$^{-1}$ and $340$ cm$^{-1}$, respectively, belonging to the in-plane and out-of-plane atomic vibration of S atoms which are in complete agreement to the experimental study of Ref. \cite{pi2018temperature}. On the other hand, the E$_\text{g}^1$ and A$_{\text{1g}}^1$ peaks belonging to Tl$_2$O largely preserve their positions albeit with a much smaller intensity of E$_\text{g}^1$ compared to its pristine counterpart. While there can be many factors affecting the Raman spectra (such as structural changes or interlayer spacing), we attribute this change to the strong interlayer interactions between the constituent systems of the heterostructure.

\section{Conclusion}

We performed density functional theory calculations to examine the electronic and spintronic properties of a Tl$_2$O/PtS$_2$ heterostructure. Different lateral stackings were carefully tried before arriving at the minimum energy configuration for which the lattice vibrations from the phonon band structure confirm dynamical stability of the heterostructure. We found an unusually high binding energy of -0.38 eV in Tl$_2$O/PtS$_2$ heterostructure which shows firm attachment and the presence of strong interlayer interactions between the two materials. While the electronic band structures were essentially preserved in the heterostructure, we discovered Rashba spin-splittings with large energy ($E_R=218$ meV) and momentum offset ($k_0=0.057\,\mathrm{\AA}^{-1}$) values around the gamma point resulting in the Rashba parameter $\alpha_R=7.65\,\text{eV}\,\mathrm{\AA}$, the highest among similar 2D heterostructures. We discussed the underlying mechanism, i.e., interfacial spin-orbit effects, via the evaluation of the electron localization function, orbital rearrangements and band structure projections and the atom-projected density-of-states calculations. In particular, we shed light on the peculiar feature of band structure alignment in momentum space but separation in real space by analyzing layer-projected band structures. The effect of small biaxial tensile strain was also highlighted in maintaining the characteristic Rashba features of the heterostructure. Finally, to expedite experimental confirmation of our results, we provided the Raman spectra of pristine Tl$_2$O and Tl$_2$O/PtS$_2$ heterostructure with details of different peak positions corresponding to atomic vibrations. Owing to the importance of spin generation, detection, and manipulation in a typical spintronics device, our results provide a promising platform to harness spin degree of freedom in 2D heterostructures by employing interfacial spin-orbit effects.

\begin{acknowledgement}
We thank Knut och Alice Wallenberg foundation, Kempestiftelserna and Interreg Nord for financial support. The computations were enabled by resources provided by the Swedish
National Infrastructure for Computing (SNIC) at HPC2N and NSC partially funded by the Swedish Research Council through grant agreement no. 2018-05973.
\end{acknowledgement}

\begin{competing interests}
The Authors declare no competing financial or non-financial interests.
\end{competing interests}

\begin{data availability}
The data that support the findings of this study are available from the corresponding
author upon reasonable request.
\end{data availability}

\begin{author contribution}
S. Sattar performed the calculations, S. Sattar and J. A. Larsson analysed the results.  All authors reviewed the manuscript. 
\end{author contribution}

\bibliography{main.bib}


\end{document}


\beginsupplement

\begin{figure}[!htb]
\includegraphics[width=1.0\textwidth]{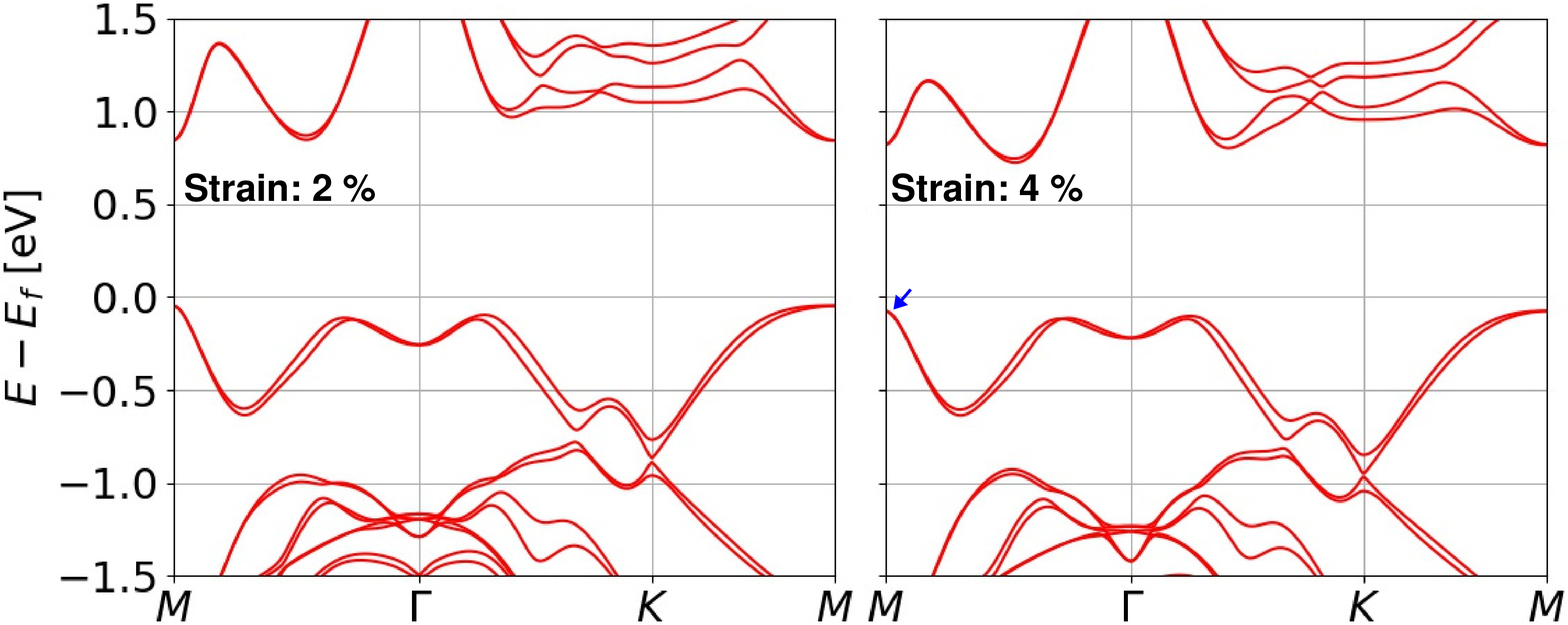}
\caption{Effect of strain on the electronic band structure of monolayer Tl$_2$O/PtS$_2$ heterostructure.}
\label{fig:figs1}
\end{figure}

\begin{figure}[!t]
\includegraphics[width=1.0\textwidth]{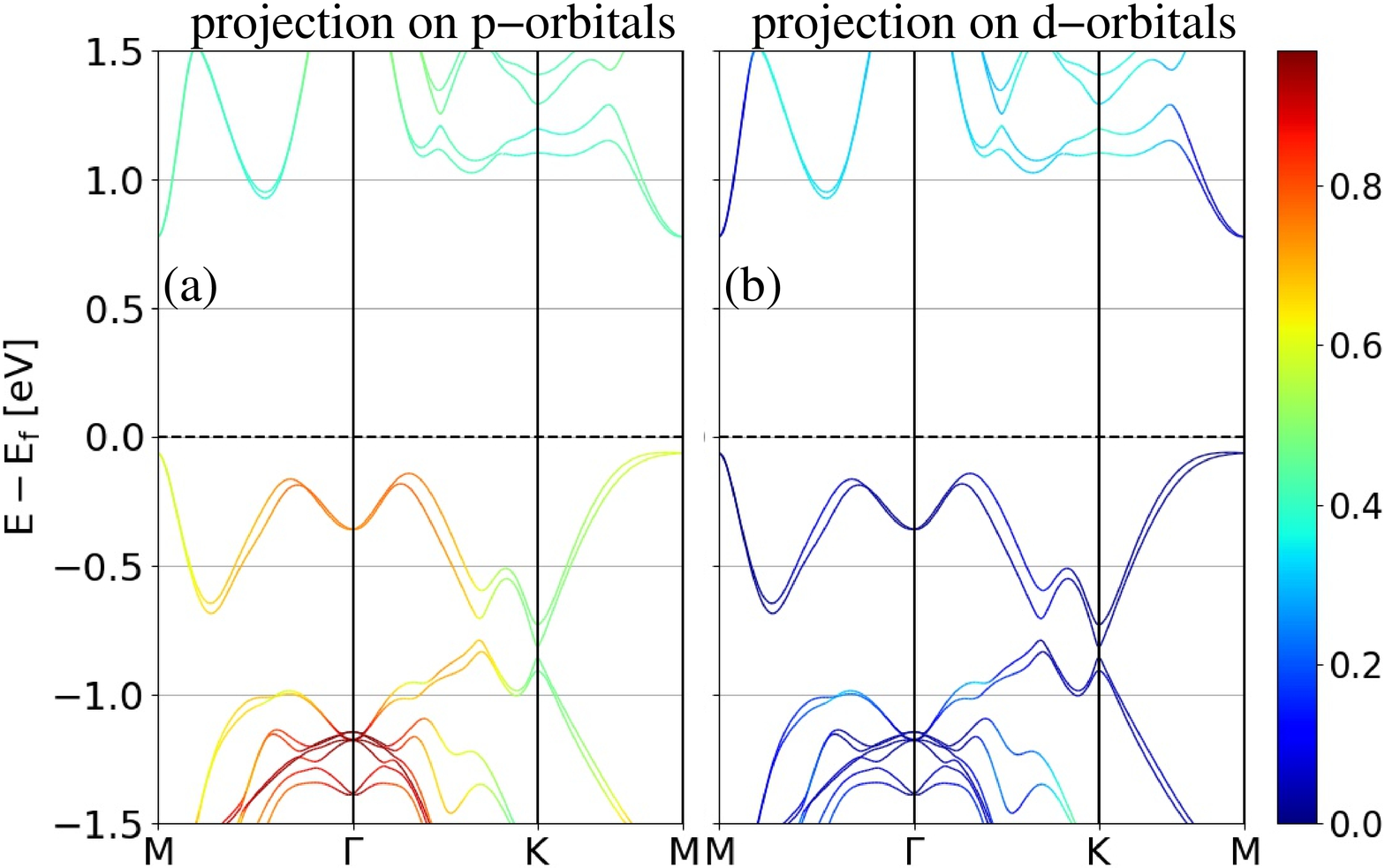}
\caption{Atomic orbital projected band structures of the Tl$_2$O/PtS$_2$ heterostructure.}
\label{fig:figs2}
\end{figure}

\clearpage